\documentclass[a4paper]{article}
\usepackage{amsmath}
\usepackage{INTERSPEECH2021}

\newcommand{\OMIT}[1]{}

\graphicspath{ {imgs/} }
\usepackage[font=scriptsize]{caption}

\usepackage{multirow}
\usepackage{tabularx}
\usepackage{array}
\usepackage{floatrow}
\usepackage{mathtools}

\usepackage{hyperref}
\newcolumntype{Z}{>{\centering\let\newline\\\arraybackslash\hspace{0pt}}X}

\makeatletter
\def\bstctlcite{\@ifnextchar[{\@bstctlcite}{\@bstctlcite[@auxout]}}
\def\@bstctlcite[#1]#2{\@bsphack
  \@for\@citeb:=#2\do{%
    \edef\@citeb{\expandafter\@firstofone\@citeb}%
    \if@filesw\immediate\write\csname #1\endcsname{\string\citation{\@citeb}}\fi}%
  \@esphack}
\makeatother

\title{Improving the expressiveness of neural vocoding with non-affine Normalizing Flows}
\name{Adam Gabryś, Yunlong Jiao, Viacheslav Klimkov, Daniel Korzekwa, Roberto Barra-Chicote}
\address{Alexa AI}
\email{\{gabrysa, jyunlong, vklimkov, korzekwa, rchicote\}@amazon.com}

\begin{document}
\bstctlcite{IEEEexample:BSTcontrol}
\maketitle

\begin{abstract}
This paper proposes a general enhancement to the Normalizing Flows (NF) used in neural vocoding.
As a case study, we improve expressive speech vocoding with a revamped Parallel Wavenet (PW).
Specifically, we propose to extend the affine transformation of PW to the more expressive invertible non-affine function.
The greater expressiveness of the improved PW leads to better-perceived signal quality and naturalness in the waveform reconstruction and text-to-speech (TTS) tasks.
We evaluate the model across different speaking styles on a multi-speaker, multi-lingual dataset.
In the waveform reconstruction task, the proposed model closes the naturalness and signal quality gap from the original PW to recordings by 10\%, and from other state-of-the-art neural vocoding systems by more than 60\%.
We also demonstrate improvements in objective metrics on the evaluation test set with L2 Spectral Distance and Cross-Entropy reduced by 3\% and 6‰ comparing to the affine PW.
Furthermore, we extend the probability density distillation procedure proposed by the original PW paper, so that it works with any non-affine invertible and differentiable function.\footnote{Audio samples will be made available on the \href{http://www.amazon.science}{amazon.science blog}.  We would like to thank Alexis Moinet, Vatsal Aggarwal and Bartosz Putrycz for insightful research discussions. And David McHardy with Jaime Lorenzo Trueba for constructive criticism of the manuscript.}
\end{abstract}
  
\section{Introduction}
Text-to-speech (TTS) is a rapidly growing domain in artificial intelligence.
TTS systems attract more attention every year as people use them for Voice Assistants, education, gaming, and much more.
High quality and low latency systems are necessary to satisfy TTS customer needs.
Most state-of-the-art Neural TTS systems address the problem of speech generation in two steps.
The first step focuses on generating a low resolution intermediate speech representation \cite{Wang2017Tacotron, Shen2018Natural, vasquez2019melnet}.
The second step concentrates on transforming this acoustic representation into a high-fidelity high-quality acoustic waveform.
Commonly, the model that refers to the second step is called the vocoder.

The state-of-the-art vocoders are generative Deep Neural Networks (DNN).
The two most prominent classes of generative models are sequential and parallel architectures.
Typically, sequential models achieve state-of-the-art results in audio, computer-vision, and textual domains \cite{pmlr-v119-jun20a}.
In neural vocoding one of the best quality sequential architectures is WaveNet \cite{Oord2016WaveNet}.
It generates waveforms autoregressively using a stack of dilated causal convolutions.
However, due to a high number of computations per sample, and the high temporal resolution of the speech signal \cite{cox2009itu} it is not suited to real-time applications.
More efficient sequential models were later proposed \cite{Kalchbrenner2018Efficient, valin2019lpcnet} that are an order of magnitude faster than WaveNet.
Nevertheless, these models are sequential, so computations cannot be easily parallelized to fully utilize modern Deep Learning ASICs, or GPUs.

The above limitation has driven most of the recent research in neural vocoding towards parallel models.
The two widely used neural vocoding parallel architectures involve Generative Adversarial Networks (GAN) \cite{Yamamoto2020Parallel, Kumar2019MelGAN, kong2020hifi, mustafa2021stylemelgan} and Normalizing Flows (NF) \cite{Oord2018Parallel, Prenger2019Waveglow, Ping2019ClariNet, kim2019flowavenet, waveffjord, kim2020wavenode}.
The generator part of a GAN can typically be any function appropriate for transforming some random inputs into synthetic outputs.
GAN-based vocoders enjoy great flexibility of architectural design and hence fast parallel generation.
The state-of-the-art adversarial-based vocoders produce high-quality, natural-sounding speech but suffer from occasional audio glitches.
These artifacts can significantly reduce the subjective score of such models \cite{jiao2021universal}.
This is a common problem related to the performance of GANs in generalizing to unseen data \cite{arora2017generalization, wu2019generalization}.
NF provides a general framework for defining probability distributions over continuous random variables.
NF takes a base distribution and transforms it to the target probability density with sequential invertible and differentiable transformations.
Normalizing Flows are compelling, in the context of vocoding, due to their efficient parallel synthesis procedure \cite{Oord2018Parallel}, and great generalization \cite{jiao2021universal}.

In this work, we focus on neural speech vocoding with Normalizing Flows (NF).
The majority of NF used in neural vocoding implement a sequence of transformations as affine functions \cite{Oord2018Parallel, Prenger2019Waveglow, Ping2019ClariNet, kim2019flowavenet}.
This type of transformation is known to be limited in its density modeling power \cite{muller2019neural, decao2019block, HuangNAF, ho2019flow++}.
In practice, we found that it adversely impacts perceived naturalness and signal quality, especially in vocoding scenarios involving highly expressive speech

The contributions of this work are:
1) We change the inexpressive affine flow transformation of PW \cite{Oord2018Parallel} to a more expressive, non-affine function;
2) We extend the probability density distillation \cite{hinton2015distilling} procedure proposed by the original PW  \cite{Oord2018Parallel}, so that it works with any non-affine invertible and differentiable function; and
3) We perform a detailed evaluation of the model across different speaking styles on a multi-speaker, multi-lingual dataset. We demonstrate that our network is qualitatively and quantitively preferred over the original PW in the waveform reconstruction and TTS tasks;
\section{Normalizing Flows \& Related Work} \label{sec: flows}
NF transforms some ${D}$-dimensional real vector of continuous random variables $\mathbf{u}$ into another ${D}$-dimensional real vector of continuous random variables $\mathbf{x}$.
Usually $\mathbf{u}$ is sampled from a simple base distribution (for example Logistic) $p_u(\mathbf{u})$. 
In the vocoding task $\mathbf{x}$ corresponds to audio signal that follows a probability density $p_x(\mathbf{x})$.
Conceptually, we can outline two blocks in the NF.
One is the transformation function $T$, which has to be invertible and differentiable.
The other is the conditioner neural network $c$ that predicts the parametrization $\mathbf{h}$ for the transformation $T$.
\begin{equation}
\mathbf{x} = T( \mathbf{u}; \mathbf{h}) \qquad \mathbf{u} = T^{-1}(\mathbf{x}; \mathbf{h})  \qquad  \mathbf{h}=c(\mathbf{u})
\end{equation}
Given the invertible and differentiable nature of $T$, the density of $\mathbf{x}$ is well-defined and can be obtainable by a change of variables:
\begin{equation}
p_x( \mathbf{x}) = p_u( \mathbf{u}) |det J_T( \mathbf{u})|^{-1}
\label{eq:nf_density}
\end{equation}
The Jacobian $ J_T(\mathbf{u})$ is $D \text{ x } D$ matrix of all partial derivatives of $T$ over $\mathbf{u}$.
In this section, we discuss the merits and limitations of different NF architectures and outline our model design.

There are two major paradigms of training NF.
One paradigm is to fit NF to the data with Maximum Likelihood Estimation (MLE) \cite{Prenger2019Waveglow,kim2019flowavenet, waveffjord, kim2020wavenode}.
In practice, it means that the model computes $T^{-1}$ during training and $T$ during the synthesis.
Another paradigm assumes that we can evaluate the target data density, and we aim to train a NF to minimize the divergence loss.
Commonly this is done with knowledge distillation \cite{hinton2015distilling}, where the data density is estimated through a teacher network \cite{Oord2018Parallel, Ping2019ClariNet}.
A notable example of this training in the context of vocoding is the use of a high-quality Wavenet \cite{Oord2016WaveNet} to train a NF-based PW  \cite{Oord2018Parallel}.
This paradigm for training and synthesis requires only the forward transformation $T$.
In both paradigms, to train the model, we have to compute the Jacobian determinant, which typically costs $O(D^3)$.
However, in many practical applications, we can reduce this complexity.
An autoregressive conditioner network has the Jacobian that is a lower triangular matrix with determinant computable in $O(D)$ \cite{Oord2018Parallel,Prenger2019Waveglow,Ping2019ClariNet,kim2019flowavenet,Kingma2016Improving, kingma2018glow}.
It is shown \cite{HuangNAF, jaini2019sumofsquares} that under the assumption of enough capacity and data, an autoregressive conditioner with non-linear transformations can approximate any continuous distribution with any desired precision - a property called universal approximation.
NF using such a conditioner can parallelize the forward transformation computation, but its inverse is sequential.
This poses a challenge for the MLE paradigm training due to the high temporal resolution of speech data.
Coupling layers are a common workaround for this problem   \cite{Prenger2019Waveglow, kingma2018glow, durkan2019neural}.
Such an architecture allows efficient computation of both forward and inverse transformations.
However, it may limit the expressivity of NF since a significant number of dimensions are left unchanged at each flow layer \cite{papamakarios2019normalizing}.
Because of the above argumentation, in this work, we decided to use ParallelWavenet \cite{Oord2018Parallel} which is a fully-autoregressive model trained with knowledge distillation that does not require any computation of the transformation inverse.

The NF transformation has to be invertible and differentiable.
The most straightforward and common design choice is to implement the transformation as an affine function \cite{Oord2018Parallel,Prenger2019Waveglow,Ping2019ClariNet,kim2019flowavenet,Kingma2016Improving, kingma2018glow}.
Such a design is attractive because of its simplicity and analytically tractability.
However, the drawback of such a transformation is its limited expressivity.
Specifically, the output of NF belongs to the same distribution family as its base.
In some cases, this might negatively affect the capture of multimodal target distributions \cite{decao2019block,HuangNAF,  ho2019flow++}.
To overcome this limitation, the transformation might be implement as a composition or the weighted sum of monotonically increasing activation functions \cite{decao2019block,HuangNAF,  ho2019flow++}, the integral of some positive function \cite{jaini2019sumofsquares}, or a spline of analytically invertible monotonic functions \cite{muller2019neural,durkan2019neural,durkan2019cubicspline}.
All above transformations are Universal Approximators \cite{HuangNAF, jaini2019sumofsquares}.
Another idea is to use constrained residual functions \cite{behrmann2019invertible, chen2020residual}. 
Unfortunately, for these methods, we either cannot efficiently compute the determinant of the Jacobian or the function has limited expressivity \cite{papamakarios2019normalizing}.
Finally, we might also construct the flow by defining an ordinary differential equation (ODE) that describes the evolution of NF in time instead of a finite sequence of transformations \cite{waveffjord, kim2020wavenode}.
According to recent surveys \cite{Kobyzev_2020}, Normalizing Flows with finite composition of non-affine transformations outperform other flow-based methods.
Considering the above pros and cons, we decide to enhance PW with a composition of monotonically increasing non-affine activation functions inspired by Flow++ \cite{ho2019flow++}.

\section{Model description} \label{sec: model}
\subsection{Parallel Wavenet} \label{sec:original_pw}
The original Parallel Wavenet \cite{Oord2018Parallel} uses conditional Inverse Autoregressive Flows \cite{Kingma2016Improving} to shift and scale base a logistic distribution to model the probability density of audio waveforms.
The procedure is as follows.
First, to generate condtioning features $\mathbf{m}$, we pass the sequence of Mel-spectrograms\footnote{Original PW uses linguistic features instead of acoustic signal representation such as Mel-spectrogram.} through transposed convolutions that upsample it to match the audio wavefrom of length $D$.
Then, we sample a $D$-long sequence of noise from Logistic distribution $\mathbf{u} \sim \mathcal{L}(0, 1)$.
We aim to model the audio waveform $\mathbf{x}$ with affine transformations that shift and scale the input noise $\mathbf{u}$.
To predict the transformation scales $\pmb{\alpha}$ and shifts $ \pmb{\beta}$ we use 
residual gated causal convolutions (RGCNN) \cite{Oord2018Parallel, oord2016conditional}.
RGCNN takes as an input conditioning $\mathbf{m}$ and a sequence of the noise $\mathbf{u}$.
For the $t$-th time step the predicted audio sample $x_t$ is:
\begin{equation}
\label{eq:aff_pw_transform}
\begin{gathered}
x_t = \alpha_t \cdot u_t + \beta_t  \\
\alpha_t, \beta_t = \text{RGCNN}(\mathbf{u_{<t}}, \mathbf{m})
\end{gathered}
\end{equation}
Multiple instances of such Flows are stacked on top of each other to increase the expressivity of NF.

Parallel Wavenet \cite{Oord2018Parallel} is trained with probability density distillation.
It is defined as KLD loss $D_{KL}$ between the teacher $P_T$ given student predictions, and student $P_S$ distributions.
In general KLD can be defined as the difference between Cross Entropy $H(P_S, P_T)$ and Entropy $H(P_S)$ terms:
\begin{equation}
D_{KL}(P_S||P_T) = H(P_S, P_T) - H(P_S)
\end{equation}
In the original Parallel Wavenet student distribution follows a Logistic function, and we can compute student Entropy analytically.
\begin{equation}
\begin{gathered}
H(P_S) = \mathbb{E}_{u \sim \mathcal{L}(0, 1)}\biggl[\sum_{t=1}^{D}\ln(\beta_t)\biggr] +2D \\
\end{gathered}
\end{equation}
The Cross Entropy term is computed via Monte Carlo approximation.
For every sample $x$ we draw from the student $p_S$, we compute all $p_T(x_t|x_{<t})$ with the teacher, and then evaluate $H(p_S(x_t|x_{<t}),p_T(x_t|x_{<t}))$.
\begin{equation}
H(P_S, P_T) = \sum_{t=1}^{D}   \mathbb{E}_{p_S(\mathbf{x_{<t}})} H(p_S(x_t|\mathbf{x_{<t}}), p_T(x_t|\mathbf{x_{<t}}))
\label{eq:cross-entropy}
\end{equation}
Sampling from the student does not require passing noise through the NF.
In a single forward pass, we cache parametrization for the Logistic Distribution, and in Monte Carlo sampling, we apply the reparametrization trick \cite{kingma2014autoencoding}.

\subsection{Non-affine transformation} \label{sec: nafftransform}
In the original PW, a student can only output a uni-modal Logistic distribution per time step, and therefore is not able to reconstruct a multi-modal mixture of Logistics (MoL) of a Wavenet teacher \cite{Oord2016WaveNet}.
To overcome this limitation, we propose to extend the affine transformation of the original PW to a non-affine function.
Inspired by the Flow++ \cite{ho2019flow++}, we implement transformation $T$ as a cumulative distribution function (CDF) for a mixture of $N$ logistics (MoL) followed by an inverse sigmoid (logit) $\sigma^{-1}$ and an affine transformation.
Such transformation is invertible and differentiable. 
The MoL CDF domain is (0, 1), so a logit of it always exists.
Also, both MoL CDF and logit functions are monotonically increasing, though invertible.
Logistics are parameterized by shifts $\pmb{\mu}$, and scales $\pmb{s}$ that are combined with mixing proportions $\pmb{\pi}$.
The output of the logit is scaled by $\alpha$ and shifted by $\beta$.
For the $t$-th time step, the predicted audio sample $x_t$ is:
\begin{equation}
\label{eq:naff_pw_transform}
\begin{gathered}
x_t = \sigma^{-1}(\textrm{MoLCDF}(u_t; \pmb{\pi_t}, \pmb{\mu_t}, \mathbf{s_t})) \cdot \alpha_t + \beta_t \\
\alpha_t, \beta_t,  \pmb{\pi_t}, \pmb{\mu_t}, \pmb{s_t} = \textrm{RGCNN}(\mathbf{u_{<t}}, \mathbf{m})
\end{gathered}
\end{equation}
Comparing to the affine function (equation \ref{eq:aff_pw_transform}), such a transformation is non-affine and can induce multimodality \cite{HuangNAF}.
The computation of the Jacobian of the transformation is straightforward since the derivative of MoLCDF is the MoL probability density function (PDF).
We also know the derivatives of logit and affine functions:
\begin{equation}
\begin{aligned}
\frac{\partial x_t}{\partial u_t} = \exp( & \alpha_t + \textrm{MoLPDF}(u_t, \pmb{\pi_t}, \pmb{\mu_t}, \mathbf{s_t}) \\
                                                          &- \ln( \textrm{MoLCDF}(u_t, \pmb{\pi_t}, \pmb{\mu_t}, \mathbf{s_t})) \\
                                                          &- \ln(1 -  \textrm{MoLCDF}(u_t, \pmb{\pi_t}, \pmb{\mu_t}, \mathbf{s_t}))) \\
\end{aligned}
\label{eq:jacob}
\end{equation}
MoLCDF and MoLPDF are defined as:
\begin{equation}
\begin{gathered}
\textrm{MoLCDF}(u_t, \pmb{\pi_t}, \pmb{\mu_t}, \mathbf{s_t}) \coloneqq \sum_{i=1}^{N} \pi_{ti} \cdot \sigma(z_{ti}) \\
\textrm{MoLPDF}(\cdots) \coloneqq  \sum_{i=1}^{N} \pi_{ti} \cdot (z_{ti} - \ln(s_{ti}) - 2 \ln(1 + e^{z_{ti}})) \\
\textrm{where} \quad z_{ti} = \frac{u_t - \mu_{ti}}{s_{ti}}
\end{gathered}
\end{equation}

\subsection{Generic and efficient training procedure} \label{sec:training_proc}
The student distribution with non-affine transformation is no longer a uni-modal Logistic.
Because of that, we have to adapt the KLD computation in the training procedure.
As in the original PW, described in section \ref{sec:original_pw}, we use Monte Carlo approximation to estimate KLD. 
However, in addition to computing predicted samples $\mathbf{x}$ with the reparametrization trick \cite{kingma2014autoencoding}, which is required to estimate Cross-Entropy with equation \ref{eq:cross-entropy}, we also compute a Jacobian determinant with equation \ref{eq:jacob}.
We do that with cached transformation parameters for every noise sample sequence.
The Jacobian allows us to evaluate the final student density $p_S$ with equation \ref{eq:nf_density}.
We use this to estimate Entropy with:
\begin{equation}
\begin{aligned}
H(P_S) &= \mathbb{E}_{u \sim \mathcal{L}(0, 1)}\biggl[\sum_{t=1}^{D}-\ln(p_S(x_t|\mathbf{u_{<t}}))\biggr] \\
            &= \mathbb{E}_{u \sim \mathcal{L}(0, 1)}\biggl[\sum_{t=1}^{D}-\ln(p_u( u_t) |det J_T( u_t)|^{-1})\biggr]
\end{aligned}
\end{equation}

\section{Experiments} \label{sec: experiments}
\subsection{Experimental setup}
\subsubsection{Training \& Evaluation datasets.}
All models used in evaluations were trained on internal studio-quality recordings.
The dataset used for training contains 22 male and 52 female voices speaking 27 languages and dialects in 10 different speaking styles.
Data were balanced so that there are approximately 3000 utterances per speaker. 
The dataset we use has a diverse range of speech vocoding scenarios and is motivated by our assumption that the non-affine transformation improves modeling more expressive distributions. 

For evaluation, we extracted Mel-spectrograms from the original studio-quality recordings.
The dataset contains 2700 recorded sentences covering 20 languages with 26 speakers in 10 different speaking styles. The dataset is balanced. There are at least 100 recordings per style and 50 per speaker.
For the subset of 1950 utterances, we also generated Mel-spectrograms in a given style from the text with Tacotron-2 like systems. \cite{Shen2018Natural}.
\subsubsection{Evaluation setup}
We run two types of evaluations.
First is the subjective evaluation that compares the affine PW with the proposed non-affine model.
We execute it as the preference tests of the TTS and waveform reconstruction tasks between the two systems.
To quantify differences we run a MUltiple Stimuli with Hidden Reference and Anchor (MUSHRA) \cite{Recommendation2001BS} evaluation of the waveform reconstruction task.
Apart from the two PW systems it also includes original recordings and two other state-of-the-art neural vocoding models: WaveGlow \cite{Prenger2019Waveglow} and ParallelWaveGAN \cite{Yamamoto2020Parallel}.
Second is the objective evaluation, which compares affine and non-affine models in terms of Cross-Entropy between the teacher and student and L2 Spectral Distance of the reconstructed waveform. 

For hypothesis testing with the objective metrics and MUSHRA we use a two-tailed t-test.
For the preference test, we evaluate a one-tailed hypothesis with the Binomial test.
We consider the difference between systems to be statistically significant if the $p-value$ is lower than $0.05.$
All subjective tests are executed on the Clickworker platform \cite{Clickworker}.
Each of the screens is assessed by 40 native listeners in the preference tests and 20 in the MUSHRA tests.

\subsubsection{Training setup \& Model details}
All of the PW \cite{Oord2018Parallel} models were distilled from a high-quality Wavenet teacher \cite{Oord2016WaveNet}.
The teacher network uses 24 layers with 4 dilation doubling cycles, 128 RGCNN channels, kernel size  3,  and output distribution of 10 MoLs.
For the student architecture, we use 2 flows with 10 and 30 RGCNN layers with 128 channels and dilation reset every 10 layers\footnote{Original PW \cite{Oord2018Parallel} uses 4 flows with 10, 10, 10, 30 RGCNN layers utilizing 64 channels. We use different hyperparameters that in our case improve the student quality.}.
Non-affine PW uses a mixture of 10 logistics in the transformation MoLCDF.
Both models were trained on Mel-spectrogram conditioning corresponding to short audio clips, with the Adam optimizer \cite{Kingma2015Adam} and a constant learning rate $10^{-4}$ for 4 million iterations with KLD and power loss \cite{Oord2018Parallel}.
The teacher uses a batch size of 64 and audio clips of 0.3625s duration, while the student uses 16 and 0.85s respectively.
WaveGlow \cite{Prenger2019Waveglow} and ParallelWaveGAN \cite{Yamamoto2020Parallel} models were trained using open-source implementations\footnote{\tiny \url{github.com/NVIDIA/waveglow}\quad \url{github.com/kan-bayashi/ParallelWaveGAN}}.

\subsection{Objective evaluation}\label{objective_metrics}
To objectively evaluate the differences between the non-affine and original PW, we propose two benchmarking metrics.
The first is the L2 Spectral Distance between the original waveform $\mathbf{x}$ and a waveform reconstructed from a Mel-spectrogram $\hat{\mathbf{x}}$. 
We transform the signal to spectrum using the short-time Fourier transform (STFT) with hop-size of 256 samples and 1024 bins. The metric is computed as ${\Vert |STFT(\mathbf{x})| - |STFT(\hat{\mathbf{x}})| \Vert}_2$.
The second is the Cross-Entropy between student and teacher under the student distribution.
The latter can be interpreted as a negative log-likelihood, which is a common metric used to evaluate NF \cite{Kobyzev_2020}.
It is computed with a Monte Carlo approximation as outlined in equation \ref{eq:cross-entropy}.
In Table \ref{tab:objective_metrics_testset}, we present average results of objective metrics attained by affine and non-affine transformations across different speaking styles.
The non-affine PW outperforms the original model on every style.
The results are statistically significant for all styles except News Briefing.
For more subjectively expressive styles, like Singing, we observe a bigger relative difference between affine and non-affine models than for less expressive ones, like Neutral.
\begin{table}[h]
\setlength\tabcolsep{2.5pt}
\scriptsize
\centering
\begin{tabular}{|l|c|c|c|c|c|c|}
\hline
{} & \multicolumn{3}{|c|}{L2 Spectral Distance} &  \multicolumn{3}{|c|}{Cross Entropy} \\
\cline{2-7}
\multicolumn{1}{|c|}{Style}  &  affine & non-affine & RD &   affine & non-affine & RD \\
\hline
News briefing     &     0.083{\tiny±0.007} & \textbf{0.078}{\tiny±0.001} & -6.3\% & 4.96{\tiny±0.07} & \textbf{4.92}{\tiny±0.07} & -8.7‰ \\
Singing           &     0.073{\tiny±0.005} & \textbf{0.068}{\tiny±0.005} & -6.1\% & 4.97{\tiny±0.07} & \textbf{4.95}{\tiny±0.07} & -4.3‰ \\
Spelling          &     0.051{\tiny±0.003} & \textbf{0.049}{\tiny±0.003} & -4.2\% & 4.35{\tiny±0.07} & \textbf{4.32}{\tiny±0.07} & -6.2‰ \\
Disc Jockey        &     0.058{\tiny±0.003} & \textbf{0.055}{\tiny±0.003} & -4.0\% & 4.71{\tiny±0.06} & \textbf{4.68}{\tiny±0.06} & -7.2‰ \\
Jokes             &     0.055{\tiny±0.003} & \textbf{0.053}{\tiny±0.003} & -3.5\% & 4.27{\tiny±0.06} & \textbf{4.25}{\tiny±0.05} & -5.7‰ \\
Long Form                &     0.056{\tiny±0.004} & \textbf{0.055}{\tiny±0.004} & -3.1\% & 4.67{\tiny±0.08} & \textbf{4.64}{\tiny±0.08} & -6.9‰ \\
Emotional         &     0.072{\tiny±0.005} & \textbf{0.070}{\tiny±0.005} & -3.0\% & 4.88{\tiny±0.04} & \textbf{4.86}{\tiny±0.04} & -5.6‰ \\
Whispering        &     0.314{\tiny±0.008} & \textbf{0.305}{\tiny±0.007} & -2.9\% & 5.56{\tiny±0.06} & \textbf{5.55}{\tiny±0.06} & -2.4‰ \\
Conversational    &     0.078{\tiny±0.005} & \textbf{0.077}{\tiny±0.005} & -2.1\% & 5.02{\tiny±0.05} & \textbf{4.98}{\tiny±0.05} & -7.4‰ \\
Neutral           &     0.066{\tiny±0.003} & \textbf{0.065}{\tiny±0.003} & -2.0\% & 4.62{\tiny±0.02} & \textbf{4.60}{\tiny±0.02} & -6.6‰ \\\hline
Overall           &     0.089{\tiny±0.003} & \textbf{0.086}{\tiny±0.003} & -2.8\% & 4.76{\tiny±0.02} & \textbf{4.73}{\tiny±0.02} & -6.1‰ \\
\hline
\end{tabular}
\caption{Average objective metrics with confidence interval of 95\% computed on the test-set. Lower (better) numbers that are different with statistical signficance ($p-val < 0.05$, two-tailed t-test) are in bold. Results are sorted by relative difference (RD) between affine and non-affine transformation on L2 Spectral Distance.} 
\label{tab:objective_metrics_testset}
\end{table}
\vspace{-0.5cm}%
\subsection{Subjective evaluation}\label{perceptual_evaluation}
To understand the subjective preference of naive listeners between the proposed non-affine model, the original PW, and other state-of-the-art neural vocoding systems WaveGlow \cite{Prenger2019Waveglow} and ParallelWaveGAN \cite{Yamamoto2020Parallel}, we run three perceptual evaluations.

To quantify differences between all of the systems we evaluate the waveform reconstruction task with MUSHRA test.
We ask listeners to rate the voices in terms of their naturalness, paying attention to the quality of the audio signal and articulation clarity.
100 means the most natural and highest audio quality speech;
0 means the least natural and lowest audio quality speech.
Overall the non-affine transformation outperforms affine ParallelWavenet with statistical significance $p-val < 0.05$.
The non-affine transformation closes the gap from the original PW to recordings by $~10\%$, and from other state-of-the-art neural vocoding systems by more than 60\%.
Non-affine PW achieves $95.35\%$ Relative MUSHRA, when  compared  to  recordings, while affine PW has $94.83\%$.
Detailed results are reported in Table \ref{tab:oracle_internal_testset}.

\begin{table}[h]
\scriptsize
\centering
\begin{tabular}{|l|c|c|c|c|c|c|c|c|c|c|c|}
\hline
{Style} & \multicolumn{1}{|c|}{Rec.}& \multicolumn{1}{|c|}{NA-PW} & \multicolumn{1}{|c|}{A-PW} & \multicolumn{1}{|c|}{PWG} & \multicolumn{1}{|c|}{WG} \\
\hline
Singing                   &  73.59  &  \textbf{64.41} &  63.93 &  49.04 &  50.79   \\
Spelling                  &  72.39  &  \textbf{69.34}* & 68.00 & 63.48  & 59.30  \\
Jokes                     &  71.10  &  \textbf{68.89}* &  67.56 &  64.18 &  55.04  \\
Neutral                   &  70.73  &  \textbf{67.63}* &  66.89 &  61.71 &  53.52  \\
News briefing             &  70.00  &  68.04 &  \textbf{68.17} &  64.81 &  61.97   \\
Disc Jockey                &  68.64  &  \textbf{66.91}  &  66.59 &  63.56 & 61.85   \\
Conversational            &  66.96  &  \textbf{67.00} &  66.81 &  63.34 & 61.38  \\
Emotional                 &  66.78  &  65.56 & \textbf{ 66.34}* &  62.81 &  61.46  \\
Long Form          &  66.42  &  65.90 &  \textbf{66.49} &  63.57 &  62.84 \\
Whispering                &  61.97  &  \textbf{54.07} &  54.00 &  34.64 &  43.60   \\
\hline
Overall  & 68.99 & \textbf{65.78}* & 65.42 & 59.01 & 55.38  \\
\hline
\end{tabular}
\caption{The MUSHRA evaluation of naturalness and signal-quality. Systems are: affine (A-PW), non-affine (NA-PW) Parallel Wavenet, WaveGlow (WG), ParallelWaveGAN (PWG), and recordings (Rec.). The highest (best) scores are in bold. '*' means that the difference between A-PW and NA-PW is statistically significant ($p-val < 0.05$, two-tailed t-test). Results are sorted by score of recordings.}
\label{tab:oracle_internal_testset}
\vspace{-0.7cm}%
\end{table}
To get a more sensitive subjective preference between affine and non-affine systems, we evaluate the waveform reconstruction task with a simple preference test.
Listeners can select either preference towards one of the systems or no-preference.
In case of no-preference, we split the votes equally between both systems.
Overall preference towards the non-affine model is confirmed with statistical significance $p-val < 0.05$.
Results across different speaking styles are statistically significant for Spelling, Singing, News briefing, and Jokes.
For these, the non-affine PW is preferred for all except News briefing. Results are presented in Table  \ref{tab:aff_vs_naff_preference}.

To understand if non-affine improvements hold for Mel-spectrograms synthesized with a Tacotron-2 like NTTS system we run an additional preference test.
The non-affine system overall outperforms the affine one with statistical significance $p-val < 0.05$.
Results for specific styles are mostly statistically insignificant, except Whisper which is better with the non-affine model.
Results are reported in the Table \ref{tab:aff_vs_naff_preference}
\begin{table}[h]
\scriptsize
\centering
\begin{tabular}{|l|c|c|c|c|c|}
\hline
{} & \multicolumn{2}{|c|}{Reconstruction} & \multicolumn{2}{|c|}{TTS} \\
\cline{2-5}
{Style} & non-affine & affine & non-affine & affine  \\
\hline	 	 	
Spelling                  &  \textbf{51.92}*   &  48.08 &   \textbf{50.87} 	& 	49.13 \\
Singing                   &  \textbf{51.81}*   &  48.19 &   - &  -   \\
Jokes                     &  \textbf{51.65}*  &  48.35 &  49.45 	& 	\textbf{50.55}  \\
Conversational            &  \textbf{50.42}  &  49.58 &   \textbf{ 50.16} 	& 	49.84  \\
Emotional                 &  \textbf{50.35}  &  49.65 &    \textbf{50.33} 	& 	49.67 \\
Whispering                &  \textbf{50.28}  &  49.72 &    \textbf{51.16}* 	& 	48.84 \\
Long Form         &  \textbf{50.16}  &  49.84 &    \textbf{50.03} 	& 	49.97 \\
Neutral                   &  \textbf{50.12}  &  49.88 &    \textbf{50.42} 	& 	49.58    \\
Disc Jockey                &  49.08  &  \textbf{50.92} &    49.77 	&	\textbf{50.23}     \\
News briefing             &  49.01  &  \textbf{50.99} &   \textbf{50.45} 	& 	49.55  \\
\hline
Overall & \textbf{50.29}*  &  49.71 & \textbf{50.42}* 	&	49.58 \\
\hline
\end{tabular}
\caption{The preference tests between affine and non-affine PW in the task of waveform reconstruction and TTS. Numbers correspond to the percent of votes towards the given system. No-preference votes are split between the two systems equally. The preferred system scores are in bold. '*' means that results are statistically significant ($p-val < 0.05$, one-tailed binomial test). Results are sorted by preference in reconstruction task.} 
\label{tab:aff_vs_naff_preference}
\end{table}
\vspace{-0.7cm}%
\section{Conclusions \& Future Work} \label{sec: discuss}
In this work, we presented a general improvement to the probability density modeling power of Normalizing Flows (NF) used in neural vocoding.
We enhanced Parallel Wavenet (PW) with a monotonically increasing non-affine activation function.
The proposed model closed the naturalness and signal quality gap from the original PW to recordings by 10\%, and from other state-of-the-art neural vocoding systems by more than 60\%.
It also reduced the L2 Spectral Distance and the Cross-Entropy computed on the multi-speaker, multi-lingual test set by 3\% and 6‰ compared to the affine PW.
For more expressive styles like Singing, we observe more improvements than for less expressive ones. 

This work motivates several possible directions for further research.
1) Non-affine NF might significantly reduce the memory footprint and the number of floating-point operations in the neural vocoding.
It is reported in other domains  \cite{decao2019block, HuangNAF} that non-affine models achieve the same or better quality than the affine ones with a much lower number of layers.
2) The non-affine transformation might simplify the complex teacher selection process for knowledge distillation models. 3) Finally, it might help to improve vocoding quality of other NF-based neural vocoding architectures. 
\bibliographystyle{IEEEtran}
\bibliography{mybib}

\begin{thebibliography}{10}
\def\url#1{}
\csname url@samestyle\endcsname
\providecommand{\newblock}{\relax}
\providecommand{\bibinfo}[2]{#2}
\providecommand{\BIBentrySTDinterwordspacing}{\spaceskip=0pt\relax}
\providecommand{\BIBentryALTinterwordstretchfactor}{4}
\providecommand{\BIBentryALTinterwordspacing}{\spaceskip=\fontdimen2\font plus
\BIBentryALTinterwordstretchfactor\fontdimen3\font minus
  \fontdimen4\font\relax}
\providecommand{\BIBforeignlanguage}[2]{{%
\expandafter\ifx\csname l@#1\endcsname\relax
\typeout{** WARNING: IEEEtran.bst: No hyphenation pattern has been}%
\typeout{** loaded for the language `#1'. Using the pattern for}%
\typeout{** the default language instead.}%
\else
\language=\csname l@#1\endcsname
\fi
#2}}
\providecommand{\BIBdecl}{\relax}
\BIBdecl

\bibitem{Wang2017Tacotron}
\BIBentryALTinterwordspacing
Y.~Wang, R.~J. Skerry{-}Ryan, D.~Stanton, Y.~Wu, R.~J. Weiss, N.~Jaitly
  \emph{et~al.}, ``Tacotron: {A} fully end-to-end text-to-speech synthesis
  model,'' \emph{CoRR}, vol. abs/1703.10135, 2017.
  \url{http://arxiv.org/abs/1703.10135}
\BIBentrySTDinterwordspacing

\bibitem{Shen2018Natural}
J.~Shen, R.~Pang, R.~J. Weiss, M.~Schuster, N.~Jaitly, Z.~Yang \emph{et~al.},
  ``Natural {TTS} synthesis by conditioning wavenet on {MEL} spectrogram
  predictions,'' in \emph{{ICASSP}}, 2018, pp. 4779--4783.

\bibitem{vasquez2019melnet}
S.~Vasquez and M.~Lewis, ``Melnet: A generative model for audio in the
  frequency domain,'' \emph{arXiv preprint arXiv:1906.01083}, 2019.

\bibitem{pmlr-v119-jun20a}
\BIBentryALTinterwordspacing
H.~Jun, R.~Child, M.~Chen, J.~Schulman, A.~Ramesh, A.~Radford \emph{et~al.},
  ``Distribution augmentation for generative modeling,'' in \emph{Proceedings
  of the 37th International Conference on Machine Learning}, ser. Proceedings
  of Machine Learning Research, H.~D. III and A.~Singh, Eds., vol. 119.\hskip
  1em plus 0.5em minus 0.4em\relax PMLR, 13--18 Jul 2020, pp. 5006--5019.
  \url{http://proceedings.mlr.press/v119/jun20a.html}
\BIBentrySTDinterwordspacing

\bibitem{Oord2016WaveNet}
A.~van~den Oord, S.~Dieleman, H.~Zen, K.~Simonyan, O.~Vinyals, A.~Graves
  \emph{et~al.}, ``Wavenet: {A} generative model for raw audio,'' in \emph{The
  9th {ISCA} Speech Synthesis Workshop}, 2016, p. 125.

\bibitem{cox2009itu}
R.~V. Cox, S.~F. D.~C. Neto, C.~Lamblin, and M.~H. Sherif, ``Itu-t coders for
  wideband, superwideband, and fullband speech communication [series
  editorial],'' \emph{IEEE Communications Magazine}, vol.~47, no.~10, pp.
  106--109, 2009.

\bibitem{Kalchbrenner2018Efficient}
\BIBentryALTinterwordspacing
N.~Kalchbrenner, E.~Elsen, K.~Simonyan, S.~Noury, N.~Casagrande, E.~Lockhart
  \emph{et~al.}, ``Efficient neural audio synthesis,'' in \emph{Proceedings of
  the 35th International Conference on Machine Learning {ICML}}, 2018, pp.
  2415--2424.  \url{http://proceedings.mlr.press/v80/kalchbrenner18a.html}
\BIBentrySTDinterwordspacing

\bibitem{valin2019lpcnet}
J.-M. Valin and J.~Skoglund, ``Lpcnet: Improving neural speech synthesis
  through linear prediction,'' in \emph{ICASSP 2019-2019 IEEE International
  Conference on Acoustics, Speech and Signal Processing (ICASSP)}.\hskip 1em
  plus 0.5em minus 0.4em\relax IEEE, 2019, pp. 5891--5895.

\bibitem{Yamamoto2020Parallel}
R.~Yamamoto, E.~Song, and J.~Kim, ``Parallel wavegan: {A} fast waveform
  generation model based on generative adversarial networks with
  multi-resolution spectrogram,'' in \emph{{ICASSP}}, 2020, pp. 6199--6203.

\bibitem{Kumar2019MelGAN}
\BIBentryALTinterwordspacing
K.~Kumar, R.~Kumar, T.~de~Boissiere, L.~Gestin, W.~Z. Teoh, J.~Sotelo
  \emph{et~al.}, ``Melgan: Generative adversarial networks for conditional
  waveform synthesis,'' in \emph{Advances in Neural Information Processing
  Systems 32}, 2019, pp. 14\,881--14\,892.
  \url{http://papers.nips.cc/paper/9629-melgan-generative-adversarial-networks-for-conditional
  -waveform-synthesis}
\BIBentrySTDinterwordspacing

\bibitem{kong2020hifi}
J.~Kong, J.~Kim, and J.~Bae, ``Hifi-gan: Generative adversarial networks for
  efficient and high fidelity speech synthesis,'' \emph{arXiv preprint
  arXiv:2010.05646}, 2020.

\bibitem{mustafa2021stylemelgan}
A.~Mustafa, N.~Pia, and G.~Fuchs, ``Stylemelgan: An efficient high-fidelity
  adversarial vocoder with temporal adaptive normalization,'' in \emph{ICASSP
  2021-2021 IEEE International Conference on Acoustics, Speech and Signal
  Processing (ICASSP)}.\hskip 1em plus 0.5em minus 0.4em\relax IEEE, 2021, pp.
  6034--6038.

\bibitem{Oord2018Parallel}
\BIBentryALTinterwordspacing
A.~van~den Oord, Y.~Li, I.~Babuschkin, K.~Simonyan, O.~Vinyals, K.~Kavukcuoglu
  \emph{et~al.}, ``Parallel wavenet: Fast high-fidelity speech synthesis,'' in
  \emph{Proceedings of the 35th International Conference on Machine Learning
  {ICML}}, 2018, pp. 3915--3923.
  \url{http://proceedings.mlr.press/v80/oord18a.html}
\BIBentrySTDinterwordspacing

\bibitem{Prenger2019Waveglow}
R.~Prenger, R.~Valle, and B.~Catanzaro, ``Waveglow: {A} flow-based generative
  network for speech synthesis,'' in \emph{{ICASSP}}, 2019, pp. 3617--3621.

\bibitem{Ping2019ClariNet}
\BIBentryALTinterwordspacing
W.~Ping, K.~Peng, and J.~Chen, ``Clarinet: Parallel wave generation in
  end-to-end text-to-speech,'' in \emph{International Conference on Learning
  Representations (ICLR)}, 2019.
  \url{https://openreview.net/forum?id=HklY120cYm}
\BIBentrySTDinterwordspacing

\bibitem{kim2019flowavenet}
S.~Kim, S.-G. Lee, J.~Song, J.~Kim, and S.~Yoon, ``Flowavenet: A generative
  flow for raw audio,'' \emph{arXiv preprint arXiv:1811.02155}, 2018.

\bibitem{waveffjord}
N.~{Wu} and Z.~{Ling}, ``Waveffjord: Ffjord-based vocoder for statistical
  parametric speech synthesis,'' in \emph{ICASSP 2020 - 2020 IEEE International
  Conference on Acoustics, Speech and Signal Processing (ICASSP)}, 2020, pp.
  7214--7218.

\bibitem{kim2020wavenode}
H.~Kim, H.~Lee, W.~H. Kang, S.~J. Cheon, B.~J. Choi, and N.~S. Kim, ``Wavenode:
  A continuous normalizing flow for speech synthesis,'' \emph{arXiv preprint
  arXiv:2006.04598}, 2020.

\bibitem{jiao2021universal}
Y.~Jiao, A.~Gabry{\'s}, G.~Tinchev, B.~Putrycz, D.~Korzekwa, and V.~Klimkov,
  ``Universal neural vocoding with parallel wavenet,'' in \emph{ICASSP
  2021-2021 IEEE International Conference on Acoustics, Speech and Signal
  Processing (ICASSP)}.\hskip 1em plus 0.5em minus 0.4em\relax IEEE, 2021, pp.
  6044--6048.

\bibitem{arora2017generalization}
S.~Arora, R.~Ge, Y.~Liang, T.~Ma, and Y.~Zhang, ``Generalization and
  equilibrium in generative adversarial nets (gans),'' in \emph{International
  Conference on Machine Learning}.\hskip 1em plus 0.5em minus 0.4em\relax PMLR,
  2017, pp. 224--232.

\bibitem{wu2019generalization}
B.~Wu, S.~Zhao, C.~Chen, H.~Xu, L.~Wang, X.~Zhang \emph{et~al.},
  ``Generalization in generative adversarial networks: A novel perspective from
  privacy protection,'' \emph{arXiv preprint arXiv:1908.07882}, 2019.

\bibitem{muller2019neural}
T.~M{\"u}ller, B.~McWilliams, F.~Rousselle, M.~Gross, and J.~Nov{\'a}k,
  ``Neural importance sampling,'' \emph{ACM Transactions on Graphics (TOG)},
  vol.~38, no.~5, pp. 1--19, 2019.

\bibitem{decao2019block}
N.~De~Cao, W.~Aziz, and I.~Titov, ``Block neural autoregressive flow,'' in
  \emph{Uncertainty in Artificial Intelligence}.\hskip 1em plus 0.5em minus
  0.4em\relax PMLR, 2020, pp. 1263--1273.

\bibitem{HuangNAF}
\BIBentryALTinterwordspacing
C.~Huang, D.~Krueger, A.~Lacoste, and A.~C. Courville, ``Neural autoregressive
  flows,'' \emph{CoRR}, vol. abs/1804.00779, 2018.
  \url{http://arxiv.org/abs/1804.00779}
\BIBentrySTDinterwordspacing

\bibitem{ho2019flow++}
J.~Ho, X.~Chen, A.~Srinivas, Y.~Duan, and P.~Abbeel, ``Flow++: Improving
  flow-based generative models with variational dequantization and architecture
  design,'' in \emph{International Conference on Machine Learning}.\hskip 1em
  plus 0.5em minus 0.4em\relax PMLR, 2019, pp. 2722--2730.

\bibitem{hinton2015distilling}
G.~Hinton, O.~Vinyals, and J.~Dean, ``Distilling the knowledge in a neural
  network,'' \emph{arXiv preprint arXiv:1503.02531}, 2015.

\bibitem{Kingma2016Improving}
\BIBentryALTinterwordspacing
D.~P. Kingma, T.~Salimans, and M.~Welling, ``Improving variational inference
  with inverse autoregressive flow,'' \emph{CoRR}, vol. abs/1606.04934, 2016.
  \url{http://arxiv.org/abs/1606.04934}
\BIBentrySTDinterwordspacing

\bibitem{kingma2018glow}
D.~P. Kingma and P.~Dhariwal, ``Glow: Generative flow with invertible 1x1
  convolutions,'' \emph{arXiv preprint arXiv:1807.03039}, 2018.

\bibitem{jaini2019sumofsquares}
P.~Jaini, K.~A. Selby, and Y.~Yu, ``Sum-of-squares polynomial flow,'' in
  \emph{International Conference on Machine Learning}.\hskip 1em plus 0.5em
  minus 0.4em\relax PMLR, 2019, pp. 3009--3018.

\bibitem{durkan2019neural}
C.~Durkan, A.~Bekasov, I.~Murray, and G.~Papamakarios, ``Neural spline flows,''
  \emph{arXiv preprint arXiv:1906.04032}, 2019.

\bibitem{papamakarios2019normalizing}
G.~Papamakarios, E.~Nalisnick, D.~J. Rezende, S.~Mohamed, and
  B.~Lakshminarayanan, ``Normalizing flows for probabilistic modeling and
  inference,'' \emph{arXiv preprint arXiv:1912.02762}, 2019.

\bibitem{durkan2019cubicspline}
C.~Durkan, A.~Bekasov, I.~Murray, and G.~Papamakarios, ``Cubic-spline flows,''
  \emph{arXiv preprint arXiv:1906.02145}, 2019.

\bibitem{behrmann2019invertible}
J.~Behrmann, W.~Grathwohl, R.~T. Chen, D.~Duvenaud, and J.-H. Jacobsen,
  ``Invertible residual networks,'' in \emph{International Conference on
  Machine Learning}.\hskip 1em plus 0.5em minus 0.4em\relax PMLR, 2019, pp.
  573--582.

\bibitem{chen2020residual}
R.~T. Chen, J.~Behrmann, D.~Duvenaud, and J.-H. Jacobsen, ``Residual flows for
  invertible generative modeling,'' \emph{arXiv preprint arXiv:1906.02735},
  2019.

\bibitem{Kobyzev_2020}
\BIBentryALTinterwordspacing
I.~Kobyzev, S.~Prince, and M.~Brubaker, ``Normalizing flows: An introduction
  and review of current methods,'' \emph{IEEE Transactions on Pattern Analysis
  and Machine Intelligence}, p. 1–1, 2020.
  \url{http://dx.doi.org/10.1109/TPAMI.2020.2992934}
\BIBentrySTDinterwordspacing

\bibitem{oord2016conditional}
A.~v.~d. Oord, N.~Kalchbrenner, O.~Vinyals, L.~Espeholt, A.~Graves, and
  K.~Kavukcuoglu, ``Conditional image generation with pixelcnn decoders,''
  \emph{arXiv preprint arXiv:1606.05328}, 2016.

\bibitem{kingma2014autoencoding}
D.~P. Kingma and M.~Welling, ``Auto-encoding variational bayes,'' \emph{arXiv
  preprint arXiv:1312.6114}, 2013.

\bibitem{Recommendation2001BS}
I.~Recommendation, ``{BS}. 1534-1. method for the subjective assessment of
  intermediate sound quality ({MUSHRA}),'' \emph{International
  Telecommunications Union, Geneva}, 2001.

\bibitem{Clickworker}
Clickworker,
  ``https://www.clickworker.com/machine-learning-ai-artificial-intelligence/,''
  August, 2020.

\bibitem{Kingma2015Adam}
\BIBentryALTinterwordspacing
D.~P. Kingma and J.~Ba, ``Adam: {A} method for stochastic optimization,'' in
  \emph{3rd International Conference on Learning Representations {ICLR}}, 2015.
   \url{http://arxiv.org/abs/1412.6980}
\BIBentrySTDinterwordspacing

\end{thebibliography}
\end{document}